\documentclass[aps,pre,twocolumn,groupedaddress]{revtex4-1}

\usepackage{graphicx}
\usepackage{amssymb}
\usepackage{amsmath}

\newcommand{\xv}{{\bf x}}

\begin{document}

\title{Anatomy of triply-periodic network assemblies:  Characterizing skeletal and inter-domain surface geometry of block copolymer gyroids}

\author{Ishan Prasad}
\affiliation{Department of Chemical Engineering, University of Massachusetts, Amherst, Massachusetts 01003, USA} 

\author{Hiroshi Jinnai}
\affiliation{Institute of Multidisciplinary Research for Advanced Materials (IMRAM), Tohoku University, Katahira, Aoba-ku, Sendai, 980-8577, Japan}

\author{Rong-Ming Ho}
\affiliation{Department of Chemical Engineering, National Tsing Hua University, Hsinchu 30013, Taiwan}

\author{Edwin L. Thomas}
\affiliation{Department of Material Science and Nano Engineering, Rice University, Houston, TX 77005, USA}

\author{Gregory M. Grason}
\affiliation{Department of Polymer Science and Engineering, University of Massachusetts, Amherst, Massachusetts 01003, USA}
\email[]{grason@mail.pse.umass.edu}

\date{\today}

\begin{abstract}
Triply-periodic networks (TPNs), like the well-known gyroid and diamond network phases, abound in soft matter assemblies, from block copolymers (BCPs), lyotropic liquid crystals and surfactants to functional architectures in biology.  While TPNs are, in reality, volume-filling patterns of spatially-varying molecular composition, physical and structural models most often reduce their structure to lower-dimensional geometric objects: the {\it 2D interfaces} between chemical domains; and the {\it 1D skeletons} that thread through inter-connected, tubular domains.  These lower-dimensional structures provide a useful basis of comparison to idealized geometries based on triply-periodic minimal, or constant-mean curvature surfaces, and shed important light on the spatially heterogeneous packing of molecular constituents that form the networks.  Here, we propose a simple, efficient and flexible method to extract a 1D skeleton from 3D volume composition data of self-assembled networks.  We apply this method to both self-consistent field theory predictions as well as experimental electron microtomography reconstructions of the double-gyroid phase of an ABA triblock copolymer.  We further demonstrate how the analysis of 1D skeleton, 2D inter-domain surfaces, and combinations therefore, provide physical and structural insight into TPNs, across multiple length scales.  Specifically, we propose and compare simple measures of {\it network chirality} as well as {\it domain thickness}, and analyze their spatial and statistical distributions in both ideal (theoretical) and non-ideal (experimental) double gyroid assemblies.
\end{abstract}


\maketitle



\section{\label{sec:intro}Introduction
}

Triply-periodic network assemblies, called here {\it triply-periodic networks} (TPNs) for short, are some of the most geometrically complex ordered structures to form in soft matter \cite{Luzzati1968, Scriven1976, Zeng2005, hyde1996language, Hyde2012}.  Like the most commonly reported examples, cubic gyroid or diamond network phases, TPNs are composed of continuously connected and interpenetrating nanoscale domains of alternating chemical composition.  These so-called {\it polycontinuous} architectures make TPNs attractive for a variety of functional hybrid materials applications, including mechanically-robust ion transport media \cite{goto2009development}, photonic crystals \cite{maldovan2002photonic, urbas2002bicontinuous, Michielsen2003, Saranathan2010, Vignolini2012, Salvatore2013, Saba2011} and plasmonic metamaterials\cite{Oh2013, Oh2015, Hur2011, lee201425th, Dolan2015}.

\begin{figure*}
\centering
\includegraphics[width=0.95\textwidth]{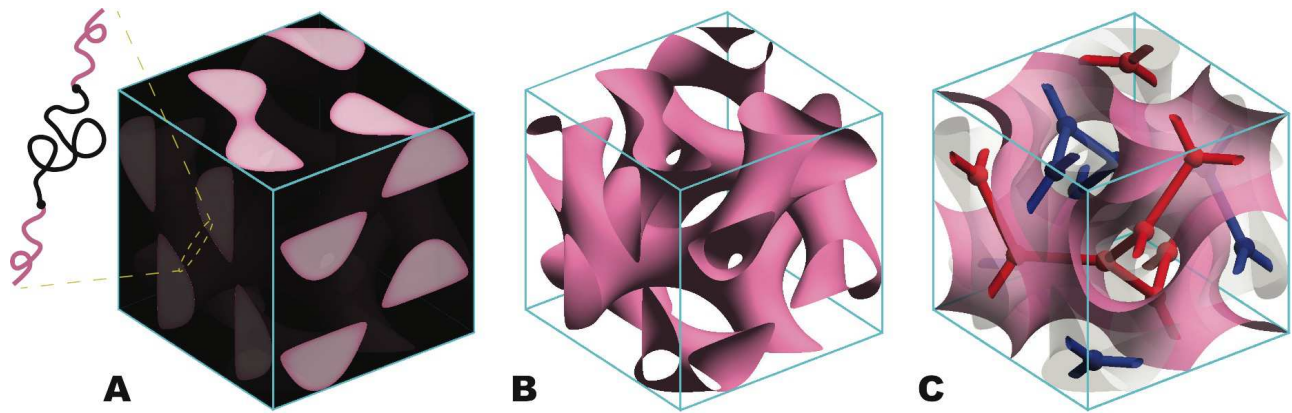}
\caption{\label{fig:1} A unit cell of the double gyroid morphology obtained from self consistent field theory calculation of an ABA triblock copolymer with $\chi N=120$, $f_A=0.32$. (A) represents the composite morphology (volume data); the matrix phase (block B) is shown in black, while the minority double gyroid domain is in pink. In (B), only the IMDS separating the double gyroid phase from the matrix is shown, the IMDS is computed from the isosurface at $\phi_A\approx0.49$ where the volume contained in the minority domain equals the fraction of the double gyroid forming S-domain ($f_S=0.32$) of the SIS triblock copolymer from Jinnai {\it et al.} \cite{jinnai2000}. (C) shows the 1D skeletal graphs associated with both disconnected single gyroid network domains and the gyroid minimal surface (in pink) separating the two networks. Translucent white jacket over each graph represents the IMDS.}
\end{figure*}

Despite their intricate structure, TPNs are observed in a large variety of soft matter systems -- including block copolymers (BCPs) \cite{Bates2005, Hajduk1994}, surfactants \cite{Fontell1990, Hyde2003, Sorenson2011}, and lyotropic liquid crystals \cite{Luzzati1968}, and biological assemblies \cite{Saranathan2010, Schroder-Turk2011, Michielsen2008} -- implying that the principles which underlie their formation are rather generic.  Indeed there is a long history that relates the thermodynamics of TPN assembly to the geometric properties of triply-periodic minimal surfaces (TPMS), and their constant mean-curvature variants \cite{Grosse-Brauckmann2012}.  Minimal area considerations arise naturally, due to the enthalpic cost (surface energy) of unlike domain contact, sometimes called the inter-material dividing surface (IMDS) \cite{thomas1988periodic}.  But as TPN assemblies are space-filling distributions of molecules, it is also not surprising that surface geometry alone is not sufficient for understanding the relative stability of different TPN symmetries. Molecules extend from one or both sides of the IMDS, and adopt configurations that fill continuous volumes -- most often, the inverse tubular ``channels'' if not also the negatively-curved, quasi 2D matrix that separates them -- at approximately constant density.  This volume-filling structure leads to distinct geometric considerations of {\it packing frustration}, or heterogeneity in local ``thickness'' of distinct domains \cite{Duesing1997, Schroeder-Turk2007}.  

While it is clear that this domain thickness must vary from place to place on the TPN structure, and also from one TPN structure to another, a rigorous, or otherwise, generically useful measure of domain thickness is still elusive.  One approach, which we follow in this article, is to analyze the geometry of the so-called 1D {\it skeletal graph} that threads through the tubular domains of TPNs.  The notion of the skeletal graph introduced by Schoen in his constructive analysis of TPMS \cite{Schoen1970} based on the duality properties of so-called saddle-polyhedra of certain infinite period nets.  It remains an open question if there exists a definition of a unique 1D {\it skeleton} for a sufficiently general class of TPN structures, (i.e. beyond a restrictive set of high-symmetry periodic graphs), though the an intuitive notion of skeleton persists.  Loosely speaking, the skeletal graph is taken to be a 1D object that traces along the geometric center of the tubular, or labyrinthine, domains of TPN, and characterizes the gross topology of the domain interconnectivity.  Heuristically, this graph is often thought of as the {\it focal domain}, representing the set of points where surface normals of the TPMS or the IMDS intersect~\cite{kleman}, though it is clear the focal domain for generic and realistic surface geometry is, in general, not 1D.  A similar approach \cite{Schroder2003, Schroder-Turk2006} has been to characterize the skeleton in terms of the 1D intersections of the 2D {\it medial surfaces} (locus of centers of maximal radii, enclosed spheres), but again, even for idealized TPMS geometries it remains unclear when and if this object has the minimal ``topological complexity'' of the graphs defined by Schoen.  

In this article, we employ an alternative and intuitive definition of the skeleton graph, as the 1D locus of {\it maximal density} points (of the enclosed chemical species) threading through a tubular domain.   As we will show, this notion of the skeletal graph is well suited to the 3D volume data (e.g., local density or intensity values) generated both from theoretical prediction and experimental characterization of TPN assemblies .  To be clear, for the present purposes, we do not intend this to be a rigorous definition of the skeletal graph, but instead, we show that it is sufficient for simple and practical numerical analysis of TPN structures whose symmetries are already known that can be applied to both ideal (theoretical) and non-ideal (experimental) TPN assembly data.  In particular, we describe and demonstrate a simple method for numerically computing the skeletal graph of double gyroid (DG) assembly formed in ABA triblock copolymers.  We apply this method to self-consistent field (SCF) theory predictions for the equilibrium patterns of alternating monomer composition, as well as 3D electron micro-tomograms (EMTs) from experiments on polystyrene-$b$-polyisoprene-$b$-polystyrene (PS-PI-PS) BCPs.  We then demonstrate how this numerically determined skeleton, in combination with the IMDS extracted via isosurface analysis of minor-block composition or scattering intensity, can be used to quantify the statistical variation of DG network structure at various scales.  Here, we compare several measures of the tubular domain thickness based on the geometry of the skeleton and IMDS, and describe how these distinct measures provide valuable insight into the well-known ``packing frustration'' in the tubular domains of BCP networks, the {\it focal distance}, as well as the distributions of closest points between 2D IMDS and 1D skeletons.  The variation of domain thickness, even in ideal TPN structures, and its implications for thermodynamics of assembly have been the subject of previous studies \cite{Schroder2003, Schroder-Turk2006, Schroder-Turk2011}.  Our present purpose is to demonstrate a simple method to apply geometric analysis in this same spirit to nanostructured BCP networks, and to compare the relative degree of inhomogeneity intrinsic to the non-uniformity of ideal TPN structures to that measured in experimental TPN reconstructions, where combined with the effects of thermal fluctuations, molecular dispersity and inevitable artifacts from the 3D reconstruction introduce nontrivial imperfections of the extracted geometry.  In this article, we show that the commonly invoked notion of {\it focal distance}, while capturing the thickness in a mean sense, fails to provide a useful measure of the local tubular domain thickness, at least with regard to constraints of molecular packing, in both experimental and theoretical gyroid structures. 

Beyond the analysis of domain thickness, we demonstrate that this simple numerical approximation of the 1D skeletal graph allows for the quantitative characterization of other key geometrical properties of the mesoscale structure of TPNs, in particular, the local, short-range and global, long range, {\it chirality}.  Among the most commonly observed TPN structures, gyroid assemblies have the unique property of structural chirality.  In the standard double-gyroid (DG) structure, each of the single gyroid (SG) sub-domains is chiral, while the arrangement of opposite chirality SG networks maintains inversion symmetry to the DG network.  Attempts to define or measure the chirality, or instead the ``handedness'', from gyroid assemblies vary.  One standard approach focuses on helical spirals traced out along the skeletal graph along various symmetry directions \cite{saba2014absence,Vignolini2012}.  For example, when viewed along a $\langle 100\rangle$, the ``larger holes'' sit at the center of the quasi-helical spirals, representing at 8 graph edges per turn.  These large spirals have been used to assign chirality in high-resolution EMT reconstruction of SGs formed in butterfly wings  \cite{Winter2015, Saranathan2010}, although in this study the definition of ``handedness'' in fact derives from different families of screws that thread the SG with opposite handedness, specifically, the smaller $4_1$ or $3_1$ screws along $\langle 100 \rangle$ and $\langle 111 \rangle$ directions, respectively.  

The ability to decompose a single enantiomeric SG network into (interconnected) spiral families of opposite handedness poses an obvious challenge to assign a unique handedness to even an ideal gyroid network, not to mention TPN structures of non-gyroid symmetries. In this article, we propose and apply a simple measure of the {\it chirality} of the distinct tubular sub-domains (single gyroid networks), that derives from distribution of dihedral angles of their 1D skeletal graphs.  On the local scale of the 1D skeleton, dihedral approach is unique as it looks at the relationship between (any) three adjacent edges in the network to determine the sense of rotation.   This method has the advantages that provides a local measure of chirality is uniform in the SG skeleton (independent of edge or direction) and is fully generalizable to networks of any topology, ordered or disordered.  Through the introduction of {\it network chirality order parameter} $\chi_{2 \theta}$ we show that we can unambiguously and quantitatively identify the distinct left- vs. right-handed chirality of the tubular SG sub-domains of experimentally reconstructed DG phases.  Notably, this is possible from non-ideal reconstructions of nanostructured BCP DGs, without identifying the multi-domain spirals that thread through perfectly ordered gyroids and without resorting to adding a component (with high electron contrast) to label the centers of tubular domains (e.g., ref. \cite{li2014linking}).  

The remainder of this article is organized as follows.  In Sec. \ref{sec:method} we first describe the methods for extraction of 3D volume data for self-assembled DG structures from ABA triblock copolymers, by SCF theoretic predictions and EMT experimental reconstructions.  We then describe our method to numerically compute the 1D graphs of the tubular minority domains based on the 3D electron contrast and monomer composition data obtained from SCF and EMT, respectively.  In. Sec. \ref{sec:results}. we exploit the 1D graph analysis, in combination with the isosurface analysis of the 2D IMDS separating minor and major domains, to analyze for both theoretical and experimental DG structures (i) the mesoscale chirality, due to rotation along the strut, of the constituent single gyroid domains and (ii) distinct measures of domain thickness and its heterogeneity.  We conclude with a discussion of possible applications and extensions of this anatomical analysis of self-organized TPN soft matter structures.  

\section{\label{sec:method}Method}

\subsection{3D volume data for BCP double gyroid \label{subsec:3Ddata}}

Here we briefly overview the methods to generate 3D volume data for DG assemblies of ABA triblock copolymers, from both computation and experiment, which is analyzed in the subsequent sections.  

{\it Theory} -  The Polymer Self-Consistent Field (PSCF) code was used to generate predictions for composition profiles of ABA triblock copolymer forming the cubic DG phase, within the mean-field (or SCF) approximation for composition fluctuations.  Details of the PSCF method are described elsewhere \cite{Arora2016}.  Briefly, an ABA triblock chain was modeled with $f_A=0.32$ volume fraction on its A end blocks.  To describe the degree of enthalpic repulsion between styrene and isoprene blocks for the molecular weight described above, $\chi N = 120$ was chosen, where $\chi$ is the Flory-Huggins repulsion parameter between A and B segments, and $N$ is the total number of chain segments.  The equilibrium solution for DG was computed in the basis $Ia\bar{3}d$ symmetric functions (space group $230$), resulting in an equilibrium size of the cubic cell repeat of $D=3.56~N^{1/2}a$ where $a$ is the statistical segment length.  From this mean-field solution we extracted 3D volume data for the local composition of the minority, A block, $\phi_{SCF}({\bf x})$, within the 3D cubic repeat cell of the DG.  To determine the IMDS, we compute an isosurface corresponding to nearly equal A and B composition, such that the enclosed volume in A-block domains match the experimental $S$-block composition. We use the isosurface at $\phi_{SCF}({\bf x})\approx0.49$, which encloses 32\% of the sample volume in the tubular domains while the rest belongs to the matrix phase. The 3D density distribution and corresponding IMDS isosurfaces for the SCF predictions for a unit cell of the DG are shown in Fig. \ref{fig:1} A and B, respectively. The gyroid minimal surface (in pink) separating the network domains and their skeletal graphs are shown in Fig. \ref{fig:1}C. Below, we analyze a volume of $[ 2~D, 2~D, 2~D ]$, where excess volume is used to minimize influence of the boundaries of 3D volume.

{\it Experimental} -  3D EMT was used to reconstruct the bicontinuous morphology of an SIS triblock copolymer melt, as reported by Jinnai {\it et al.} \cite{jinnai2000}. In this sample, SIS chains have a number average molecular weight and polydispersity index of $8.3 \times10^4$ and 1.09, respectively,  and a volume fraction of $f_S=0.32$ of the PS block.  Details of the 3D EMT method for reconstructing BCP morphologies are reviewed elsewhere\cite{Jinnai2009}.  Briefly, the reconstruction is performed on a microtomed sample of PS-PI-PS (thickness varies from 250 to 300 nm) mounted on a tilt stage transmission electron microscope (TEM) equipped with 200KeV.  A tilt-series of TEM images are obtained at 2.1nm/pixel resolution, with the sample rotated from $\pm 60^{\circ}$ in increments of $2.5^\circ $.  Heavy atom labeling by staining the PI domain by OsO$_4$ leads to contrast between PI and PS domains.  A 3D alignment and reconstruction algorithm converts the tilt series to 3D volume data \cite{Spontak1996, Fung1996, Frank1992}, corresponding to numerical reconstruction of the scalar electron scattering contrast, $I_i$ at spatial positions ${\bf x}_n$, reconstructed on a rectilinear grid with 250, 250 and 127 pixels in the $x$, $y$ and $z$ (thickness) dimensions at a resolution of 2.1 nm/pixel.  We convert the intensity signal to an approximation of local PS composition $\phi_{EMT}({\bf x}_n)$ by assuming a simple proportionality between $I_i$ and the local density of isoprene, $\phi_{EMT}({\bf x}) = 1- I_n/I_{max}$, where $I_{max}= {\rm max}_n[I_n]$ the maximum in the contrast signal.  To construct the 2D IMDS separating the PI and PS domain, isosurfaces of constant $\phi_{EMT}({\bf x}_n)=\phi_{\rm IMDS}$ are computed (using MATLAB), with the level set parameter $\phi_{\rm IMDS}$ chosen so that enclosed volume within region of  $\phi_{EMT}({\bf x}_n)\leq \phi_{\rm IMDS}$ set to $f_S=0.32$.  For the geometric analysis described below, we analyze a roughly cubic sub-volume, with dimensions $[ 166~{\rm nm},  166~{\rm nm},  165~{\rm nm}]$, compared to the cubic repeat of the DG structure, the lattice constants, are $a =78$ nm, $b=71$ nm, and $c=74$ nm, or computational grid of approximately $2.2D$ per side.  The selected volume is highlighted (in pink) in Fig. \ref{fig:2}A, vertically centered to avoid analysis near microtome surfaces.   

\subsection{\label{subsec:graph} Skeletal graph extraction}

Here, we describe our simple method to numerically compute the 1D skeletal graphs corresponding to the two tubular minor domains of the DG (i.e., the PS, or A-block domains) from the entire 3D volume data $\phi({\bf x})$, a continuously variable scalar intensity at ${\bf x}$ (not to be confused with binary data).  The objective of the analysis is to determine spatial positions of graph vertices such that mean value of $\phi(\xv)$ along the 1D graph, $\Phi$, is maximal.  Here, we seek only a {\it local maximum} of $\Phi$, considering networks with a fixed topology corresponding to the ideal skeletal graphs of the gyroid, that is, the two enantiomeric $(10,3)-a$ networks, also called cubic $(10,3)-a$ nets, where in each vertex, or ``node'', is 3-coordinated \cite{Wells1977x}.  Defining the position of $i$-th vertex as ${\bf v}_i$, the mean value of composition is defined as
\begin{equation}
\Phi = L^{-1} \sum_{\langle ij \rangle} \int_i^j ds~ \phi({\bf x} )
\end{equation}
where $\langle ij \rangle$ indicate edges, or ``struts'', of the graph passing from $i$ to $j$, and $\int_i^j ds (\cdot) $ is line integral of the 1D line from ${\bf v}_i$ to ${\bf v}_j$ and $L$ is the total length of the graph, $L = \sum_{\langle ij \rangle} \int_i^j ds$.   Because the DG possesses two disjoint tubular, 3-fold connected nodes of minor domains, of opposite chirality, the procedure described below is performed twice to compare the two enantiomorphic graphs independently.  Here, $\Phi$ is computed by numerical integration of interpolated values of $\phi({\bf x})$ along graph edges.

\subsubsection{Initial skeletal graph}
We construct an initial graph corresponding to the one of the two enantiomeric $(10,3)-a$  nets, possessing the $I4_132$ symmetry while the other has $I4_332$ symmetry.  For a cubic repeat of distance $D$, the ``+'' graph is composed of the vertices (Wyckoff positions of $I4_132$ \cite{international1959international}),
\begin{eqnarray}
{\bf v}^+_1 = \Big(\frac{D}{8} ,\frac{D}{8},\frac{D}{8}\Big); &\!\!\!\!\!\!& {\bf v}^+_2 = \Big(\frac{D}{8},\frac{7D}{8},\frac{3D}{8}\Big); \nonumber \\
   {\bf v}^+_3 = \Big(\frac{3D}{8},\frac{D}{8},\frac{7D}{8}\Big); &\!\!\!\!\!\!&  {\bf v}^+_4= \Big(\frac{7D}{8},\frac{3D}{8},\frac{1D}{8}\Big)
\end{eqnarray}
plus the translations of these 4 vertices on a BCC lattice with primitive vectors $(\pm \frac{D}{2},\pm \frac{D}{2},\pm \frac{D}{2})$ \cite{Grobe-Brauckmann1997, Grosse-Brauckmann1997, Kaufmann2010}.  The opposite ``-'' graph is generated by inverting this graph through the cell center $(\frac{D}{2}, \frac{D}{2}, \frac{D}{2})$, or ${\bf v}^- = -{\bf v}^+ + (D,D,D)$.  Struts connecting the nodes of each of the two Wyckoff sites run parallel to the $\langle 110 \rangle$ directions and are of length $D/\sqrt{8}$.  

\begin{figure}[t]
\includegraphics[width=0.475\textwidth]{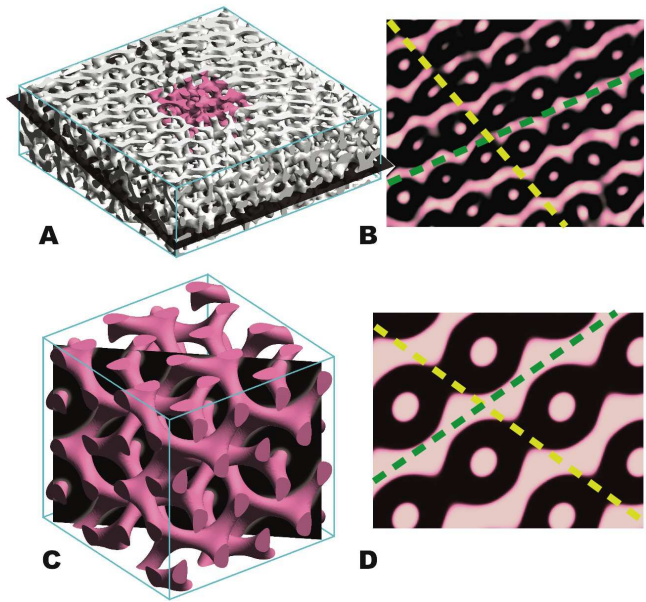}
\caption{\label{fig:2} Electron density patterns from EMT reconstruction of the SIS double gyroid morphology (light pink domains on PS) (B) on a slice plane (in black) in (A).  Comparison to the predicted composition pattern from SCF for the triblock assembly in C,D identifies this direction (normal to plane) as the $\langle110\rangle$ direction of the cubic DG cell.   The ideal SG graphs (i.e. $(10,3)-a$ nets) are aligned and scaled to match the intensity patterns in this plane: circular high intensity spots correspond to the tubular domains of one SG subnetwork passing normal to the plane; and diagonal, ``zig-zag'' patterns (highlighted by the dashed green lines in B and D) connect the struts of the other network lying in the plane.   See the Supplemental Videos for comparisons for animated comparisons along this and other symmetry axes of the DG.}
\end{figure}

\subsubsection{Graph alignment}
To determine a local fit of the graph within one of the two tubular minor domains (with appropriate chirality), the $(10,3)-a$ graph is aligned to an identifiable symmetry axis of the DG volume data. This allows for global adjustment of the graph scale ($D$) to match the periodicity of the DG structure.  For example, graphs in both EMT and SCF structures were aligned to a $\langle110\rangle$ direction of the cubic cell.  The 2D density cuts normal to this direction for both experimental and theoretical DG structures are shown in Fig \ref{fig:2}.  From this perspective the graph can be aligned with the density pattern by scaling, translating and rotating (around the $\langle110\rangle$ direction) to register the edges of the graph normal to this plane with the quasi-hexagonal pattern of circular, high density ``spots'', interspersed between diagonal, ``zig-zag'' contours (which correspond to tubular regions of the opposite SG domain), as seen in Fig. ~\ref{fig:2}B,D.  Following an initial alignment and scaling adjustment by visually superposing the graph of 2D density cuts, the value of $\Phi$ (computed for the section of the graph enclosed in the 3D data volume) is numerically optimized over rigid rotations, translations and affine scaling of $D$, leading a slight increase in the preconditioned value of the average density. At this stage, the chirality of the 1D skeleton (i.e. whether the domain is ``+'' or ``-'' SG) is guessed based on the apparent overlap of the prealigned (10,3)-a graph with high-density contours in the volume data at variable cuts along the $\langle110\rangle$.  For example, at subsequent depths, the pattern of circular spots transforms continuously to a ``zig-zag'' contour, through which the nodes and struts of the prealigned graph should be threaded, and the predominant rightward vs leftward tilt of graph edges in these layers changes depending on the chosen enantiomorphic graph. This preliminary assignment of chirality achieved in this way is verified later by comparison to fits of opposing graph to the same tubular sub-domain.  Subsequent analysis of the opposite tubular domain proceed in the same way, only with the initial graph prealigment performed at 2D cut offset by $0.25D$ in $\langle001\rangle$, such that quasi-hexagonal spots correspond to tubular ``struts'' of the this second domain.

\begin{figure}[t]
\includegraphics[width=0.475\textwidth]{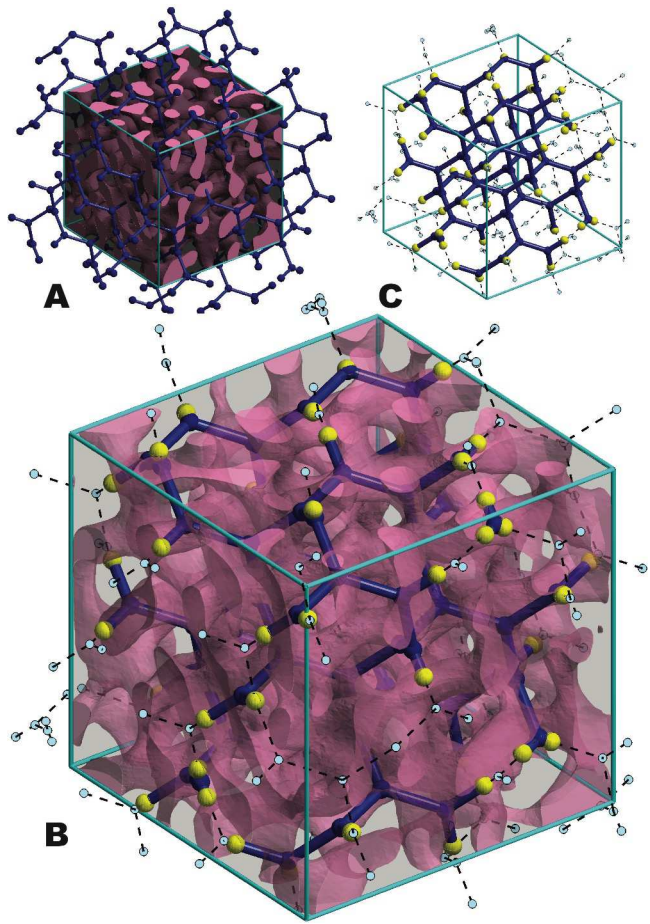}
\caption{\label{fig:3}  (A) The infinite $(10,3)-a$ graph, superposed on the EMT reconstruction of the SIS triblock morphology, after the pre-alignment step of the initialization process. Blue cylinders and spheres represent struts and vertices respectively; (B) the clipped initial graph that resides within the finite domain where $\phi(\xv)$ is defined. Additional boundary vertices where the infinite graph intersects the boundary faces of the domain are shown as yellow spheres. The exterior portions of the graph (dashed lines and light blue circles) are then ignored. (C) Shows only the clipped graph in the same sample volume.}
\end{figure}

\subsubsection{Graph clipping and vertex relaxation}
While the $\phi({\bf x})$ data is defined only over finite volume, the preconditioned gyroid graph is taken to extend over much larger spatial region, and extends beyond these nearly cuboidal volumes, denoted as ${\cal V}$.  To further relax the graph geometry, beyond the ideal $(10,3)-a$ geometry of the preconditioned graph, regions of the graph outside of ${\cal V}$ are ``clipped'' as follows.  Exterior vertices ${\bf v}_i \not\in {\cal V}$ are removed from the graph, and graph edges that protrude a boundary of ${\cal V}$ are truncated, by defining new vertex positions at the planar faces ${\cal V}$ which connect to adjoining interior vertex (see Fig.~\ref{fig:3}).  At this point the positions of graph vertices (interior and boundary) are relaxed in order to optimize $\phi$, with boundary vertices constrained to lie on the 2D plane of their respective faces of ${\cal V}$.  We use, {\it fmincon}, the constrained minimizer in MATLAB to optimize the vertex positions, to tolerance corresponding to vertex displacements smaller than $10^{-4}$ times the cell edge.  For the SCF data sets, the relaxation achieves a final mean density of $\Phi \geq 0.99$, presuming the correct enantiomer of the $(10,3)-a$ graph was used to fit the tubular domain.  For comparison, local optimum of EMT, where electron contrast $\phi_{EMT}(\xv)$ is far more diffuse than that ideal segregation predicted by SCF, reaches a maximal mean density, $\Phi \simeq 0.68$.  The overall lower contrast in the EMT data set is attributable to a host of experimental factors including distortions of the structure from microtomy and beam damage leading to misregistration of domain overlap in the projections as well as OsO$_4$ staining of PS regions~\cite{handlin1983, vanbavel2010}.

The overall lower contrast in the EMT data set is at least partially attributable to imperfect confinement OsO$_4$ staining to PS domain .

\begin{figure*}[t]
\centering
\includegraphics[width=\textwidth]{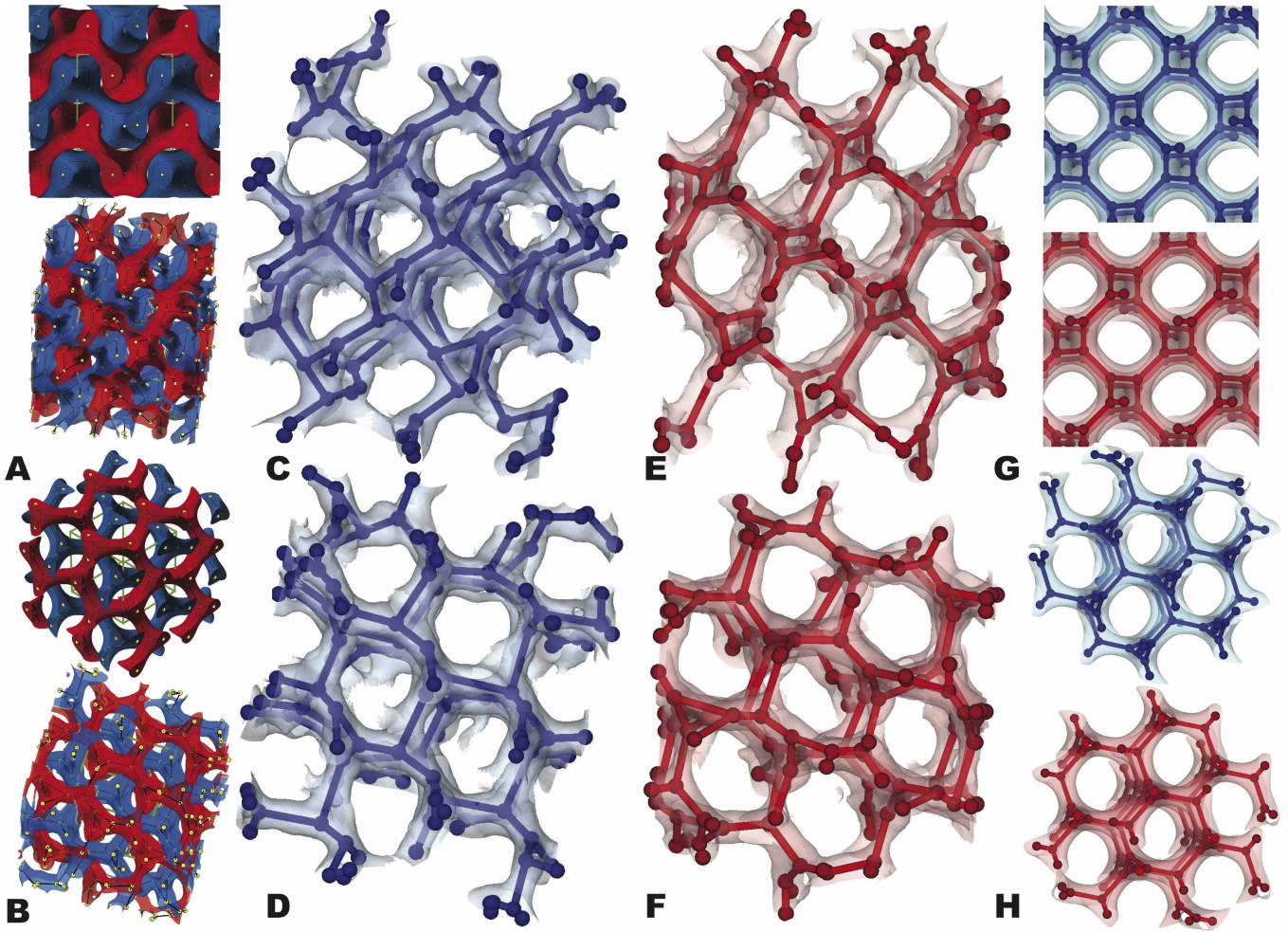}
\caption{\label{fig:4} Numerically computed skeletal graphs and their corresponding tubular domains. (A,B) Double gyroid morphologies corresponding to SCFT  calculations (above) and EMT reconstructions (below) viewed along (A) $\langle100\rangle$ and (B) $\langle111\rangle$ directions. The networks are colored according to the best fit $(10,3)-a$ net obtained from the vertex relaxation procedure. (C-F) Extracted single networks with corresponding equilibrated skeletal graphs from EMT reconstructions viewed along (C, E) $\langle100\rangle$ and (D, F) $\langle111\rangle$; (G,H) Similar views from SCFT for comparison in (G) $\langle100\rangle$ and (H) $\langle111\rangle$ directions. The red (blue) SG networks are ``+'' (``-'') according to the sense of rotation of the dihedral angle from node to node.}
\end{figure*}

Fig.~\ref{fig:4} shows the results of the skeletal graph computations from triblock DG assembly for both ``+'' and ``-'' SG tubular domains, from SCF predictions and experimental EMT reconstructions.  While skeletons of EMT data are clearly deformed from the ideal $(10,3)-a$ net geometry, the gross symmetry of SG network is clearly visible when viewed along the high symmetry directions (e.g. $\langle 100 \rangle$ and  $\langle 111 \rangle$).

\section{\label{sec:results}Results \& Discussion}

Here we use the numerically determined  skeletal graphs and IMDS, shown in Fig. \ref{fig:4}, to analyze the mesoscale geometry of tubular (minor) domains of the DG assembly.  Specifically, we illustrate how these 1D and 2D geometrical abstractions can be used to quantify i) the {\it chirality} of constituent gyroid tubular domains and ii) distinct measures of the tubular domain thickness.

\begin{figure*}[t]
\centering
\includegraphics[width=0.95\textwidth]{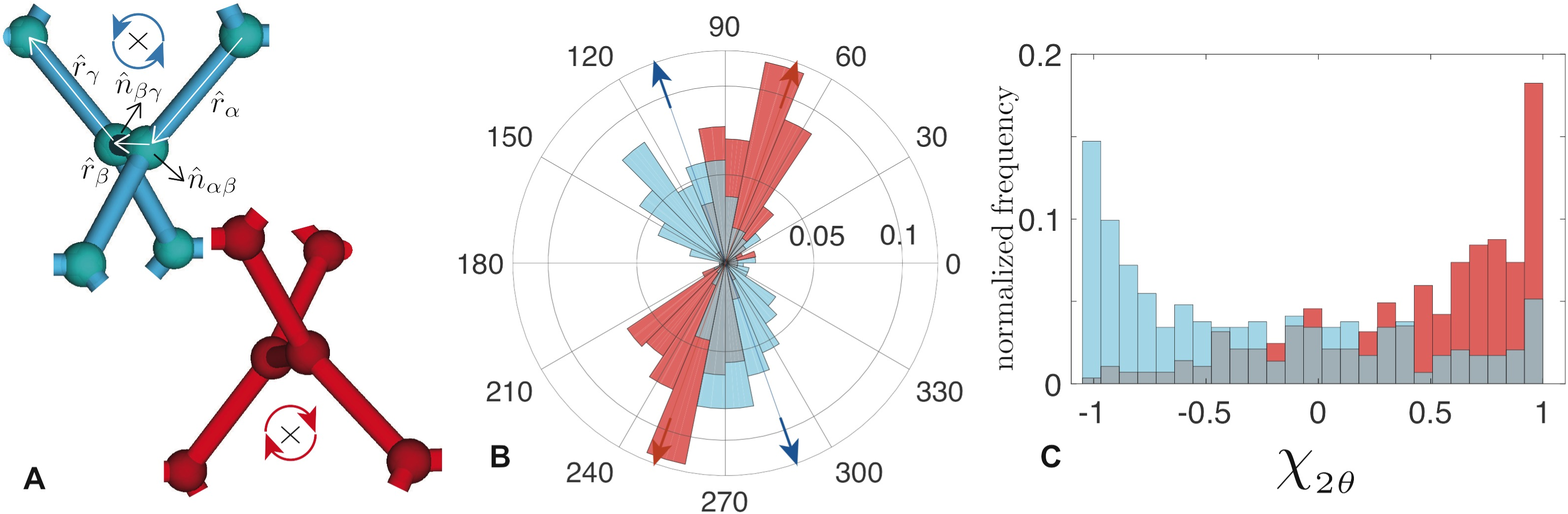}
\caption{\label{fig:5} Dihedral angle analysis to identify the local sense of rotation of each single gyroid network. Our calculation identifies the torsion angle associated with every three-strut configuration and spans the entire network. (A)  Sections of both single gyroid networks, viewed in the $\langle 110 \rangle$ direction, obtained from SCFT calculation, comprising of six nodes connected via five struts in a ``dog-bone-like'' configuration. One such torsion angle is the angle between normal vectors, $\hat{n}_{\alpha\beta}$ and $\hat{n}_{\beta\gamma}$, assigned to pair of bond vectors $\hat{r}_\alpha\hat{r}_\beta$ and $\hat{r}_\beta\hat{r}_\gamma$ respectively. (B) Comparison between the polar histograms of all torsion angles associated with the skeletal graph of ``-'' (blue) and ``+'' (red) networks. The dashed line distribution corresponds all edges of the graph while the solid distributions excludes the edges possessing vertices on the volume boundaries. Radial distance marks the probability of occurrence of the dihedral angle ($\theta$). Dark blue (dark red) arrows show the expected torsion angle for -(+) network obtained from SCFT for comparison. (C) Histograms of local chirality $\chi_{2\theta}$ (relative occurrence) evaluated on each edge (or strut) of numerically computed graphs from opposing ``+''/``-'' single gyroid domains. }
\end{figure*}

\subsection{\label{subsec:angle} Network chirality:  dihedral geometry}

We propose a simple analysis of the {\it chirality} of TPN networks (or any network) based on the distribution of the dihedral angles between adjacent edges in the graph.  For the ideal SG skeleton, this dihedral geometry is easy to visualize in terms of the planar 3-fold nodes.  As shown in Fig.~\ref{fig:5}A for the network in blue (red), the normals to planes, $\hat{n}_{\alpha\beta} $ and $\hat{n}_{\beta \gamma}$, on adjacent nodes are rotated by $\pm 70.5^\circ$, or $\mp 109.5^\circ$.  Here, sign of the dihedral, $\theta_\beta$, is determined by the right-handed rotation of normals passing along the connecting edge (e.g., in the direction of $\hat{r}_\beta$ in Fig. \ref{fig:5}A).  From the sign of $\theta_\beta$ (modulo $180^\circ$), we therefore assign positive (negative) rotation angles as right-handed (left-handed) dihedrals.   This definition has the key advantages that (a) it defines chirality at the smallest possible scale of the skeletal graph, the two-node strut,  and (b) it can be applied to any 1D graph, independent of symmetry or long-range order, to assess the spatial and statistical variation of chirality in the structure, as will be illustrated below.

To apply this analysis to realistic networks, where geometry of vertices is not necessarily planar, we define a dihedral angle for every three consecutive graph edges, with directions labeled by $\hat{r}_\alpha$, $\hat{r}_\beta$ and $\hat{r}_\gamma$, and corresponding normals, $\hat{n}_{\alpha\beta}=(\hat{r}_\alpha\times \hat{r}_\beta)/|\hat{r}_\alpha\times \hat{r}_\beta|$ and $\hat{n}_{\beta \gamma}=(\hat{r}_\beta \times \hat{r}_\gamma)/|\hat{r}_\beta\times \hat{r}_\gamma|$,
\begin{equation}
\sin \theta_\beta = (\hat{n}_{\alpha\beta}\times\hat{n}_{\beta\gamma})\cdot \hat{r}_{\beta}; \ \cos \theta_\beta = \hat{n}_{\alpha\beta}\cdot\hat{n}_{\beta\gamma} . 
\end{equation}
Applying this analysis to triplets of adjacent edges in skeletal graphs obtained from both SCF and EMT volume data (excluding boundary edges), we obtain the polar dihedral histogram in Fig. \ref{fig:5}B.  Notably, the dihedral angles of the SCF graphs are locked into the ideal gyroid values, $\pm 70.5^\circ$, or $\mp 109.5^\circ$, while the skeletons from the experimental EMT gyroids show dispersion away from this ideal geometry, owing to fluctuations in the local network chirality due to the combination of local intensity fluctuations and artifacts in the 3D reconstruction.  Both ``+'' and ``-'' networks show a systematic chiral skew, tilting respectively to the right and left of the $90^\circ$ mark in Fig. \ref{fig:5}B.  

To quantify the statistical fluctuation of local chirality, we define a network chirality order parameter,
\begin{equation}
\chi_{2\theta} \equiv \langle \sin (2 \theta_\beta) \rangle ,
\end{equation}
where $\langle \cdot \rangle$ indicates a global average over all non-boundary edges.  The symmetry of the order parameter accounts for the $\theta \to \theta + \pi$ symmetry of the dihedral angle, and a non-zero value of $\chi_{2 \theta}$ indicates a net rightward (positive) or leftward (negative) rotation of dihedrals.  Compared to the ideal gyroid graph chirality $\chi_{2\theta} ({\rm ideal}) \approx \pm 0.63$, we plot the histogram of the dihedral chirality, $\sin (2 \theta_\beta)$, of the numerically determined graphs of the ``+'' and ``-'' single gyroid graph from the EMT data in Fig. \ref{fig:5}C, which have mean values $\chi_{2\theta}^+ ({\rm EMT}) = + 0.43$ and $\chi_{2\theta}^- ({\rm EMT}) = - 0.30$, $\chi_{2\theta} ({\rm ideal})$ is marked by dark red and blue lines, respectively.  Note that the statistics only considers internal edges, with no vertices at the surface of the volume (inclusion of the boundary edges leads to a marginal decrease order parameters, $\chi_{2\theta}^+ ({\rm EMT}) = +0.36$ and $\chi_{2\theta}^- ({\rm EMT}) = - 0.26$).  While the fluctuations of the local chirality of the graph deviates the value of $\chi_{2\theta}$ somewhat from its mean value, the non-zero chirality of the EMT graphs is statistically significant, with a separation between the mean $\chi_{2\theta}$ values of ``+'' and ``-'' graphs that exceeds the variance (see Table~\ref{tbl:std}).  

We note that while dihedral rotation provides arguably the most uniform and local measure chirality possible for regular network like SG, it is possible to define other order parameters to measure chirality, for example, that weight the length of edges or nodes differently.  Indeed, it is well know that no single order parameter may be defined to properly classify structures as chiral vs. achiral because of the possibility of so-called ``false zeros" for any given pseudo-scalar measure~\cite{harris1999, efrati2014}.  Nonetheless, for the sufficiently uniform geometry of these experimental networks, $\chi_{2\theta}$ has the desirable property that when averaging over both networks simultaneously it returns a near 0 measure as expected for the net achiral symmetry of the DG.

\begin{figure*}
\centering
\includegraphics[width=0.95\textwidth]{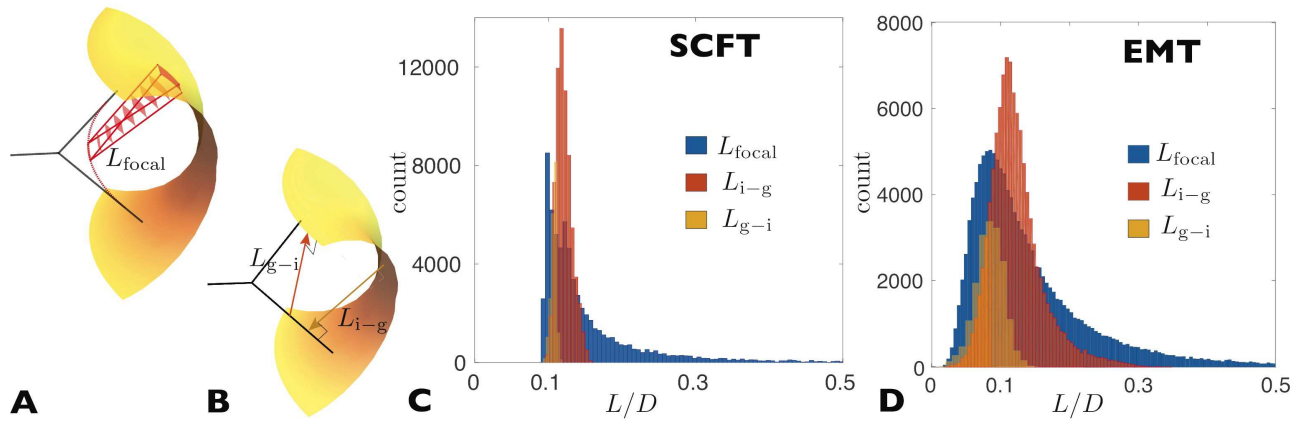}
\caption{\label{fig:6} Histograms of domain thickness in minority block network phases based on three different measures. (A) A schematic of the wedge height, $L_w$; (B) a schematic of the shortest distance between the interface and the skeletal graph measured (i) from the IMDS to the graph, $L_{\rm i-g}$ and (ii) from the graph to the IMDS, $L_{\rm g-i}$. Histogram of $L_w$, $L_{\rm i-g}$, and $L_{\rm g-i}$ from (C) SCFT calculations; (D) EMT reconstructions of ABA triblock double gyroids. }
\end{figure*}

\subsection{\label{subsec:thickness} Domain thickness:  skeleton to surface distance}

We now demonstrate how analysis of the combined 1D skeleton and 2D IMDS geometry provides a comprehensive perspective on the  {\it thickness} of tubular SG subdomains of the DG assembly.  While ``thickness'' is trivial and intuitive to define for high-symmetry domain shapes (spheres, cylinders, planar layers), achieving a unique and generically valuable metric of domain size for tubular triply periodic surface domains is a notorious vexing problem \cite{hyde1996language, Schroeder-Turk2007, Hyde2012}. The complexity of measuring tubular domain thickness in TPNs arises for two reasons.  First, given the complex domain topology, there are multiple possible measures of thickness, defined according to inequivalent geometric criteria.  And second, even in an ideal geometry of tubular domains (such as CMC surface models), intrinsic variability of domain shape leads to perturbations in any measure of thickness.  Despite the challenge to quantify this thickness variability, it has been implicated in the thermodynamic stability of some TPN structures over others in equilibrium assembly.  The variation of tubular domain thickness is associated with the ``packing frustration'' of chain-like molecules that must extend from the IMDS into the entire volume of the domain.  The observation that model geometries of gyroid assemblies can be characterized as more homogeneous in terms of one or more measure of domain size (or ``packing length'') in comparison to competitor structures, like diamond, is believed to account for lower stretching penalty for their formation \cite{matsen1997block}.

Here, we compare three distinct measures of domain thickness, $L$, and compare their statistical and structural variation for both ideal (theory) and non-ideal (experiment) DG structure:

\noindent {\it Focal distance}: $L_{\rm focal}$ - The distance, perpendicular to the IMDS, at which local normals intersect.  This measure is often considered as ``self-intersection'' distance of equally-spaced layers filled into the interior of the convex domain, although intersections of such surfaces are not strictly local.  In such a geometry, a given surface patch of the IMDS, $\Delta A_0$, sweeps out a ``wedge'' of volume  (see, Fig. \ref{fig:6}A) of corresponding to an area element $\Delta A(z)= \Delta A_0 (1+ 2 H z + K z^2)$ at normal distance $z$, where $H$ and $K$ are the mean and Gaussian curvatures of the IMDS \cite{steiner1840parallele}.  The focal distance, $L_{\rm focal}$, sometimes considered the ``wedge height'', is defined by the $\Delta A(L_{\rm focal}) = 0 $, or

\begin{equation} \label{eq:focal}
L_{\rm focal} = H/K ( 1- \sqrt{1-K/H^2} ),
\end{equation}

where we assign normals to point inward such that $H<0$ for the tubular domains.  Note that this measure is defined on surface elements of the IMDS, computed using discrete approximations of $H$ and $K$ via the triangular mesh geometry of the numerical IMDS isosurface.  Below we show the limitations of this ``surface only'' measure of domain thickness. 

\noindent {\it Surface-to-skeleton distance}: $L_{\rm i-g}$ - The closest point on the enclosed skeletal graph to a given surface element of the IMDS.  This length is computed  by finding the closest point on the 1D skeleton (each edge is discretized into 100 segments) to a given vertex of the triangular IMDS mesh, considering only the skeleton enclosed in the same tubular domain.   As shown in Fig.~\ref{fig:6}B, this closest separation from the surface to the skeleton always meets normal to the graph. 

\noindent {\it Skeleton-to-surface distance}: $L_{\rm g-i}$ - The closest point on enveloping IMDS to a given point on skeletal graph.   As shown in Fig.~\ref{fig:6}B, this closest separation from skeleton to IMDS is always perpendicular to the surface.  Unlike the other two measures of thickness which are surface distributions,  $L_{\rm g-i}$, is defined for each element on the 1D graph.    

In Fig.~\ref{fig:6} C and D, we analyze the distributions of the three domain thickness measures for theoretical and experimental triblock DG assemblies.  Note that the frequency distributions are unnormalized and correspond to different numbers of surface and discrete elements.  Note also, that while network chirality was a measure of each single gyroid tubular domain, domain thickness distribution is a measure of both SG domains.  Finally, for both $L_{\rm i-g}$ and $L_{\rm g-i}$ we exclude values corresponding to graph edges which intersect the boundary of the data volume, the closest surface of graph point of the structure lies outside of the data volume close to the boundary.  

\begin{figure*}
\centering
\includegraphics[width=\textwidth]{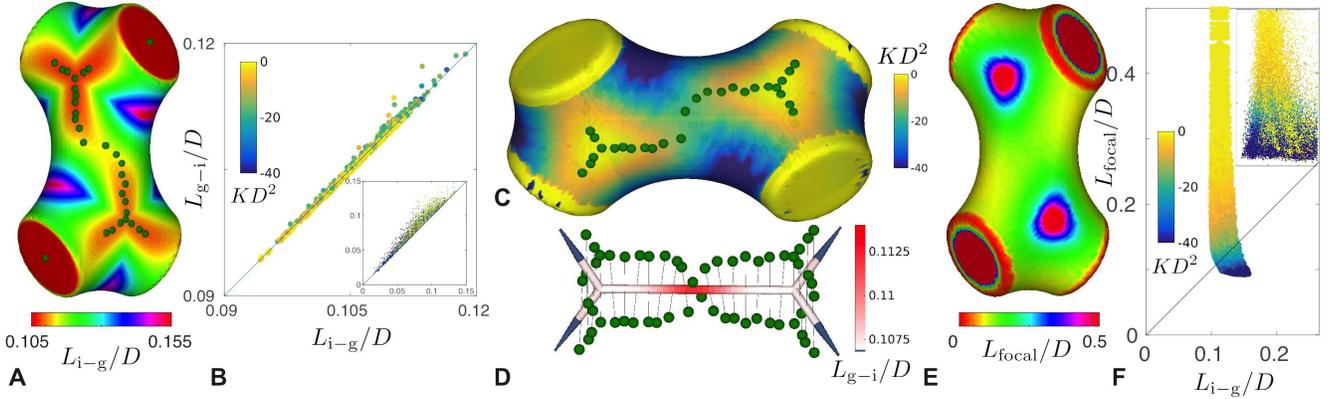}
\caption{\label{fig:7} Comparison between domain curvature and thickness from SCFT calculation. The lengths are measured in units of $D$.(A) IMDS of a two-node strut showing the colormap of the surface-to-skeleton, $L_{\rm i-g}$; (B) comparison between $L_{\rm i-g}$ and skeleton-to-surface $L_{\rm g-i}$ domain thickness measures from both SCF calculations and EMT reconstructions. Theoretical scatter points are colored according the gaussian curvature, $KD^2$, of the IMDS. The two measures $L_{\rm g-i}$ and $L_{\rm i-g}$ agree very well; inset shows the entire distribution while the main plot highlights the theoretically relevant region. Despite larger spread in data the experimental and theoretical results are consistent as shown in distributions of Fig~\ref{fig:6} (C) Strut colored by Gaussian curvature, $KD^2$, with green spheres showing that shortest skeleton-to-IMDS, $L_{\rm g-i}$ distances trace the flattest regions (vanishing gaussian curvature) of the two-node strut; (D) the extracted piece of the skeletal graph connecting two nodes colored by the $L_{\rm g-i}$ distance along the strut. Blue regions are ignored due to the artificially introduced slicing to extract the two node strut. We see that $L_{\rm g-i}$ increases with increasing magnitude of Gaussian curvature consistent with our intuition of higher stretching at higher IMDS curvature. Furthermore, the twisting feather-like pattern of the $L_{\rm g-i}$ indicates the dihedral rotation along the strut is linked to its chirality. (E) The two-node strut colored by the thickness measure dependent solely on the local curvature of the IMDS, the focal distance, $L_{\rm focal}$;  (F) Scatter plot from  SCF calculations reconstructions showing the correlation between $L_{\rm focal}$ and $L_{\rm i-g}$. Data from EMT reconstructions are plotted in the inset with the same axis limits as the main figure. Shading indicates that these thickness measures agree well (fall close the line $L_{\rm focal}=L_{\rm i-g}$) in regions of most negative Gaussian curvature of IMDS. }
\end{figure*}

First, we note that the three distributions of $L_{\rm focal}$, $L_{\rm i-g}$ and $L_{\rm g-i}$, differ in terms of both their mean values of spread, with the skeleton-to-surface distance corresponding to the shortest and narrowest distribution for both SCF and EMT structures, while the focal distance have the broadest spread.  The mean and root-mean-square-error (rmse) of chirality and each thickness measure is summarized in Table \ref{tbl:std}.  As shown, mean values of thickness measures, while different from one measure to the other, generally agree when comparing experimental reconstruction to SCF predictions for the domain structure, particularly for focal and surface-skeleton distances, while we find a roughly 20\% difference between the mean values of skeleton-to-surface distance, although this is comparable to the statistical spread distance in the EMT data.  While the analysis shows good agreement between mean (and peak) values of the distributions, the statistical spread from the EMT reconstructed domain is far greater.   This broader spread derives from the non-ideal length/angle geometry of the numerically determined skeleton and IMDS for enhanced fluctuations in $\phi({\bf x})$ in the experimental data, which are themselves attributable to the combined effect of intrinsic composition fluctuations in the morphology and microtome deformations from the tilt-series to 3D reconstruction of local intensity.  Notwithstanding this inevitable degree of disorder, we conclude on the basis of the comparison to ideal (SCF based) structures that our simple numerical abstraction of 1D skeleton and 2D IMDS geometry captures an accurate and quantitatively meaningful analysis of the thickness geometry (and its variability) for non-ideal DG structures extracted from experimental systems.

\begin{table}[h]
\small
  \caption{\ Mean and standard deviations of chiral order parameter and different measures of length (in units of $D$).}
  \label{tbl:std}
  \begin{tabular*}{0.475\textwidth}{@{\extracolsep{\fill}}lllll}
    \hline
    Measure & mean (theory) & rmse (theory) & mean (exp.) & rmse (exp.)\\
    \hline 
    $\chi_{2\theta}^-$ & -0.63 & 0 & -0.30 & 0.62 \\  \vspace{1pt}
    $\chi_{2\theta}^+$ & +0.63 & 0 & +0.43 & 0.5  \\  \vspace{1pt}
    $L_{\rm focal}$ & 0.15 & 0.08& 0.16 & 0.16 \\ \vspace{1pt}
    $L_{\rm i-g}$ & 0.12 & 0.01 & 0.13 & 0.05 \\ \vspace{1pt}
    $L_{\rm g-i}$ & 0.11 & 0.004 & 0.08 & 0.02\\
    \hline
  \end{tabular*}
\end{table}

We now discuss key differences between the average and statistical features captured by each of the domain thickness measures.  We begin by comparing the surface-to-skeleton and skeleton-to-surface thickness measures. It should be noted that these distances are geometrically distinct, in that while $L_{\rm i-g}$ measures distance normal to the skeleton, $L_{\rm g-i}$ measures distance within the minority domain normal to the IMDS.  Nevertheless, we can consider the overlap between the distributions in the following way.  For each point on the graph, ${\bf x}_g$ the closest distances provides a map to a surface element of the IMDS, ${\bf X}^*_i({\bf x}_g)$.  In Fig.~\ref{fig:7} A-B we compare the skeleton-to-surface distances $L_{\rm g-i}$ to the surface-to-skeleton of those mapped surface points (i.e. the set of ${\bf X}^*_i({\bf x}_g)$) mapped back to the closest point on the graph, though not, in general, back to the original point ${\bf x}_g$.  We find that $L_{\rm g-i} \approx L_{\rm i-g}$ for these surface elements that are among the set of closest points to the skeleton, consistent with the interpretation that the skeleton-to-surface distances represent the subset of shortest lengths among the surface-to-graph distances.  Hence, the distributions of $L_{\rm g-i}$ overlap with the low end of the distributions of $L_{\rm i-g}$. The inset shows entire distribution from both experiments and theory.

Notably, as shown in Fig.~\ref{fig:7} C the IMDS elements closest to the graph (i.e., the set of points ${\bf X}^*_i({\bf x}_g)$ shown as green) are also spatially localized to the relative flat regions, (low $K$ magnitude).  Put another way, the skeleton-to-surface distances in the tubular SG domain measure the domain thickness in precisely those regions where the focal distances from the IMDS break down as a realistic measure of the domain thickness.  As shown in Fig. ~\ref{fig:7} A, C and D, this locus of {\it closest points from the skeleton} sweeps out a spatial pattern following regions of locally low $K$ magnitude, and adopts a {\it chiral pattern} that rotates with the {\it same sense} of rotation as the dihedral angles in the SG skeletal graph.  This pattern highlights a geometric connection between local anisotropy of the domain thickness and the mesoscale geometry captured by the 1D skeleton of the SG domains.  Because these distributions of domain thickness couple to distributions of space-filling molecular configurations in the self-assembled domain, the emergent chirality of the pattern of ${\bf X}^*_i({\bf x}_g)$ on the IMDS implies a connection from mesoscale chirality of the SG domains to the chirality of molecular arrangements at the sub domain scale.  The consequences for this meso- to molecular mechanism of chirality transfer are yet to be explored for soft matter systems that form gyroid, or other triply-periodic, network assemblies.

Finally, we compare these measures to the focal distance, which unlike the previous two is determined only by local curvature data on the IMDS. We note that mean (and peak) focal distances, $\langle L_{\rm focal} \rangle$, are comparable to those distances between skeleton and surface (e.g., within the statistical variation of the $L_{\rm i-g}$ distance), such that {\it on average} the focal length provides a reasonable, if somewhat larger, measure of the distance between the IMDS and the domain ``center''.  However, even for the ideal (SCF) DG, the large-$L_{\rm focal}$ tail of this distribution extends far beyond these other distance measures, up to distances comparable to half cubic repeat of the DG.  Comparing that to the scale of the roughly tubular diameter $\approx 0.2D$ of the internode ``strut'' suggests that the focal distance in this tail region extends outside of the domain itself.  The anomalously large values of $L_{\rm focal}$ can be attributed to locally flatter regions of the IMDS where Gaussian curvature tends toward zero, leading to focal distance that diverges in the limit, according to eq. (\ref{eq:focal}).   This is a well-noted consequence of the fact that the focal domain only accounts for local geometric constraints imposed by the surface, and non-local surface constraints must be imposed to achieve a thickness measure ``internal'' to the tubular domain (e.g., as for the medial distance).  Figure~\ref{fig:7} A shows the distribution of focal distance on the IMDS surface of an SCF gyroid domain, highlighting that the larger values of $L_{\rm focal}$ correspond to nearly-flat IMDS shapes roughly coplanar with the 3-fold node.  The discrepancy between the focal length and the distance to the ``geometric center'' of the gyroid domain is illustrated by comparison of $L_{\rm focal}$ and $L_{\rm i-g}$ for surface elements of the SCF morphology.  As shown in Fig.~\ref{fig:7} B, and consistent with geometric arguments above, the overlap between these two thickness measures (i.e. $L_{\rm focal} \approx L_{\rm i-g}$) is concentrated to IMDS regions of sufficiently negative Gaussian curvature.  While fairly convenient to compute and often invoked as a proxy for ``packing length'' in TPN assemblies, the focal distance substantially exceeds realistic notions of ``domain thickness'' over large fractions of the tubular gyroid surface.  It is worth noting that the low end of the focal distance distribution is largely consistent with the distances measured using both the skeleton and IMDS, indicating that it nevertheless serves a reasonably simple, if approximate, measure of the ``narrow'' thickness regions of the structure.

\section{\label{sec:Concl} Conclusion}
In conclusion, we have demonstrated a simple numerical analysis of the 1D skeleton and 2D IMDS geometry of a nanostructured TPN assembly based on 3D intensity of local composition data.  We demonstrate a simple algorithm that optimizes the mean intensity/density along the 1D graph to determine the skeleton that threads through a tubular TPN domain.  Applying this analysis to both theoretical and experimentally reconstructed composition profiles of a self-assembled DG phase, we show that the combined geometry of the IMDS (extracted through standard isosurface methods) and the skeleton provide a far more realistic measure of the tubular domain thickness and its variations than what can be measured through local IMDS geometry itself (i.e., the focal distance).  Beyond this, we propose and analyze a generic order parameter, $\chi_{2 \theta}$, for the chirality of the network assembly, and show that this order parameter can distinguish the sense of rotation of the dihedral angle from node to node for the pair of enantiomorphic networks in the DG assembly, as well as the degree of ``statistical disorder'' of network chirality in the non-ideal experimental network.  Beyond the application to gyroids, where the ideal structure has uniform dihedral chirality, this chirality analysis is easily extended to analyze disordered networks (e.g., spinodal network morphologies~\cite{hyde1995swelling, jinnai1997curvature} or trabecular bone~\cite{ben-zvi2017} as well as other networks where dihedral chirality is non-uniform by assigning a local value of $\chi_{2 \theta}$ to each edge of the skeleton (e.g. averaged over all triplets of consecutive edges passing through that edge). According to this definition, for example, the (3-fold coordinated) skeletons of the O70 network phase~\cite{Epps2004, tyler2005}, which are formed in certain BCP systems, are overall achiral, but they could also be characterized in terms of alternating patterns of dihedral chirality (e.g., -/+) running along the orthorhombic cell.  It remains to be understood how this measure of mesoscale chirality relates to the chirality of the molecular-scale organization of the constitutes of the tubular domains, and if, for example, the value of $\chi_{2 \theta}$ for different chiral networks correlates measures of linear and non-linear chiral response (e.g., optical or acoustic modes).  

Finally, we conclude with a brief note about a limitation of the present implementation of skeleton computation, the presumption high degree of prior knowledge about the symmetry and the topology of skeletons, in this case {\it a priori} identification of the morphology as cubic DG.  Our present purposes were to develop a simple and efficient algorithm to generate a local ``best fit'' skeletal graph given an initial guess sufficiently close to what is believed to provide the global maximum mean density, and from this, to explore the numerical accuracy of this computed graph for measuring size and angular geometry of the tubular SG domains.  In general, it may be desirable to have analysis to either explore networks with large degrees of topological disorder, or instead networks where the symmetry is not known {\it a priori}.  For example, nanoscale TPN assembles are typically characterized by point link defects (e.g., broken or extra struts) as well as extended defects (grain boundaries), which clearly violate topological assumptions of ideal graph templates, and approaches such as topological thinning have been to applied to BCP networks to characterize these defect~\cite{takahashi2003}, although the such 1D networks often result in curved edges, ``excess" nodes, and contractible cycles.  Topological variations of the skeleton may be addressed in the present framework via a Monte Carlo sampling of graph topology.  For example, it would be straightforward to consider random topological moves to add, remove or join vertices, weighted according relative increase or decrease of mean intensity or density along the graph.  Of course, it remains to be determined what, if any, optimal annealing and population sampling can negotiate presumably rugged ``fitness'' landscape that are likely to characterize variable skeleton topologies for nanostructured networks that are far from a known ideal topology.  

\section*{Acknowledgements}

The authors are grateful to R. Spontak for initial discussions regarding EMT reconstructions of triblock networks and R. Kusner for useful discussions regarding TPMS.  This work was supported by the U.S. Department of Energy, Office of Science, Office of Basic Energy Sciences, under Award No. DE-SC0014549 (development and application mesogeometric analysis and SCF simulation of triblock assemblies, I. P., E. T. and G. G. ) and JSPS KAKENHI Award Nos. 16H02288 \& 16K14001 and Management Expenses Grants for National Universities Corporations from Ministry of Education, Culture, Sports, Science and Technology of Japan (3D EMT reconstruction data, H. J.).  SCF calculations were performed on computational facilities at the Massachusetts Green High Performance Computing Center.


\bibliography{Skeleton_Skin}

\end{document}